# Bottom-up realization of a type-II organic–TMD heterointerface: Pentacene on monolayer WS$_2$

Michele Capra[1,*], Christian S. Kern[2], Mira S. Arndt[1], Karl J. Schiller[1], Max Niederreiter[2], Francesco Presel[2], Iolanda Di Bernardo[5], Marco Gruenewald[3], Torsten Fritz[3], Stefan Tappertzhofen[4], Martin Sterrer[2], Peter Puschnig[2], Mirko Cinchetti[1] and Giovanni Zamborlini[1,2,*].

[1] TU Dortmund University, Department of Physics, 44227 Dortmund, Germany

[2] Institute of Physics, NAWI Graz, University of Graz, 8010 Graz, Austria

[3] Institute of Solid State Physics, Friedrich Schiller University Jena, 07743 Jena, Germany

[4] TU Dortmund University, Department of Electrical Engineering and Information Technology, 44227 Dortmund, Germany

[5] Universidad Universidad Autónoma de Madrid and IFIMAC, 28049 Madrid, Spain

*corresponding authors: michele.capra@tu-dortmund.de; giovanni.zamborlini@uni-graz.at

## Abstract

Stacked van der Waals heterostructures based on transition metal dichalcogenides (TMDs) exhibit a rich variety of exotic interfacial phenomena. Substituting one component with an organic semiconductor (OSC) enables the design of hybrid heterostructures with tunable functionalities for optoelectronic, photovoltaic, and spintronic applications.

In this work, exploiting scanning tunneling spectroscopy (STS), photoemission orbital tomography (POT) and $G_0W_0$ electronic structure calculations, we experimentally and theoretically demonstrate the self-assembly of an ordered single layer of pentacene (5A) above monolayer WS$_2$, exhibiting a type-II (staggered) band alignment in the hybrid 5A/WS$_2$ interface. Central to this result is the synthesis of extended, atomically flat WS$_2$ – an essential prerequisite for a highly ordered and electronically homogeneous OSC/TMD interface – which can only be reliably achieved via bottom-up growth, most notably molecular beam epitaxy (MBE). We realize this by leveraging Au(111) as an atomically clean and conductive sample for epitaxial growth - a necessary requirement for reliable and comparable STS/POT characterizations. The high quality of the synthesized heterostructure, together with its type-II band alignment, establishes pentacene/WS$_2$ as a model system for orbital-resolved studies of charge transfer, energy-level renormalization, and non-equilibrium interfacial processes in hybrid organic-inorganic–2D heterostructures.

## Introduction

Transition metal dichalcogenides (TMDs), a class of 2D materials where a single layer of transition metal atoms is sandwiched between two layers of chalcogens (S, Se or Te),[1,2] have attracted significant scientific interest owing to their unique electronic and optical properties. Depending on their composition, bulk TMDs can exhibit electronic properties ranging from metallic to semiconducting[1,2] to Weyl semimetals[3] (*e.g.* WTe$_2$), and display a variety of non-trivial electronic properties, including superconductivity[1,4,5] (*e.g.* NbSe$_2$), charge density wave ordering[6] (*e.g.* TaS$_2$) and spin valley splitting[1,7] (*e.g.* WS$_2$). In semiconducting group IV TMDs, it is possible to control both the size and type of the band gap via their thickness.[8,9] The band gap is inversely proportional to the TMD's thickness, and at the monolayer limit an indirect-to-direct bandgap transition also occurs[1,10,11].The plethora of these phenomena make TMDs suitable for the development of next generation electronics, photonics and optoelectronics applications[12-15], like photodetectors[16,17], solar cells[18] and transistors[19]. Moreover, stacking together different 2D materials to form van der Waals heterostructures has been shown to induce new properties that do not arise from the individual constituents of the junctions, but rather from their interface[20,21]. As an example, such interfaces can host interlayer excitons[20], where the electron and the hole are located in two separate materials (at the interface), or moiré states, resulting from twisting two different 2D materials[21].

Being based on atomically thin materials, where each layer simultaneously acts as both bulk and interface, the surface quality plays a crucial role in the performance of 2D heterojunctions. Up to now, the most common approaches for the synthesis of heterojunctions have relied on top-down methods[22], like exfoliation of bulk crystals with tape or liquid sonication, which typically results in only micrometer-sized and poor-quality interface areas. Bottom-up approaches, though more demanding form a technical point of view, enable the synthesis of highly crystalline, almost defect-free, and millimeter-scale interfaces[23,24]. Beyond structural advantages, these methods also open up unique opportunities for nanoscale functionalization, making them particularly promising for tailored molecular integration[25,26]. In particular, hybrid interfaces between TMDs and organic semiconducting (OSC) molecules are gaining interest due to the ability of molecules to efficiently tune optical, electronic and magnetic properties of 2D materials[26] combined with a seamless implementation, low cost and an almost infinite choice of organic molecules.

In this context, interfacing OSCs with TMDs holds the promise of coupling the high responsivity to external stimuli offered by OSCs with the superior charge transport of monolayer TMDs [27,28], as also supported by recent breakthroughs in the exciton-mediated optical control of proximity effects at organic/metal interfaces[29,30]. Extending this concept to semiconducting heterojunctions unlocks new opportunities for photovoltaic, optoelectronic and sensing applications. However, the performance of such hybrid systems critically depends on the relative alignment of molecular and TMD energy levels[28,30], which defines the direction and efficiency of charge transfer across the interface. Depending on the band alignment, three main heterojunction types can be identified: type-I (straddling), type-II (staggered), and type-III (broken-gap). Among those, type-II architectures are particularly attractive, as their staggered band alignment promotes spatial charge separation between the layers. Consequently, such interfaces may support the formation of long-lived interlayer excitons [27], enabling efficient energy transfer from the strongly light absorbing organic layer to the high-mobility TMD substrate[27,29,30]

In this study, we synthetize a hybrid interface comprised of single-layer WS$_2$ and a single layer of pentacene (5A) that self-assembles atop the TMD, forming a long-range ordered film. The electronic and geometric properties of the interface have been revealed with a multi-technique approach that combines scanning tunneling microscopy (STM) and spectroscopy (STS), angle-resolved photoemission spectroscopy (ARPES) and photoemission orbital tomography (POT), supported by ab-initio electronic structure calculations at the density functional theory (DFT) and the $G_0W_0$ level, respectively. Inspired by earlier works on $MoS_2$, $NbS_2$ and $VS_2$,[31-33] we demonstrate here the first successful growth of extended, atomically flat $WS_2$ single crystals using dimethyl disulfide (DMDS) as a sulfur source, an environmentally benign and non-toxic alternative to the commonly used $H_2S$.[34,35] Upon adsorption of the 5A layer, we observe the formation of a commensurate molecular superstructure characterized by type-II band alignment, in which the highest-occupied molecular orbital (HOMO) of 5A lies within the $WS_2$ band gap, while the lowest-unoccupied molecular orbital (LUMO) overlaps with the $WS_2$ conduction band. Our bottom-up approach not only ensures high crystalline quality but also provides clean, chemically accessible surfaces, crucial for the controlled assembly of molecular layers and the engineering of well-defined excitonic interfaces.

## METHODS

**Experimental.** The experiments have been performed in three different UHV chambers with a base pressure better than $2 \times 10^{-10}$ mbar.

The Au(111) sample was cleaned by repeated cycles of sputtering with 2 keV $Ar^+$ ions for 40 minutes, followed by annealing at 950 K for 30 minutes. The $WS_2$ monolayer was grown by reactive Molecular Beam Epitaxy (MBE) in a four steps process, evaporating tungsten in dimethyl disulfide (DMDS) atmosphere. In order to favor the complete sulfurization of tungsten atoms, DMDS was injected in the UHV chamber exploiting a nozzle placed as close as possible to the Au(111) surface. First, the Au(111) substrate was exposed to DMDS at 950 K for 10 minutes, with the nozzle placed at 1 mm from the surface, with a chamber back-pressure $p_{DMDS} = 2.5 \times 10^{-7}$ mbar in order to deposit sulfur atoms on the surface before deposition of tungsten. The second step consisted in the deposition of tungsten in DMDS atmosphere, chamber back-pressure of $p_{DMDS} = 3.5 \times 10^{-7}$ mbar with the nozzle placed at 5 mm from the surface, still holding the sample at 950 K. The evaporation rate was calibrated in order to achieve 1 ML of tungsten, equal to the atomic surface density of Gold, in 15 minutes. The third step consisted in a post-annealing of the grown $WS_2$ in DMDS atmosphere, using a chamber back-pressure $p_{DMDS} = 2.5 \times 10^{-7}$ mbar with the nozzle placed at 1 mm from the surface, at 950 K for 30 minutes. Finally, the fourth step consisted in a second post-annealing without DMDS, lowering the temperature from 950 K to 775 K, in order to stabilize the grown structure and desorb DMDS molecules that remained stuck to the surface during the growth. To avoid nucleation of $WS_2$ into a second layer, we deposited a nominal coverage of 0.7 ML, resulting in the exposure of some clean gold surface.

The pentacene (5A) monolayer was deposited on the as-grown $WS_2$/Au(111) by thermal evaporation from a commercial Knudsen cell (Kentax GmbH) at 450 K. The 5A evaporation rate was calibrated by means of a quartz microbalance.

X-ray Photoelectron Spectroscopy (XPS) was performed with a monochromatized Al Ka radiation source (SPECS µFocus 600) and an electron analyzer with a spectral energy resolution of 400 meV (SPECS PHOIBOS 150). XPS spectra were fitted using a pseudo-Voigt line shape, with Lorentz-Gauss weights of, respectively, 0.80-0.20, and the kinetic to binding energy conversion is given with respect to the Fermi edge of the Au(111) substrate.

Scanning tunneling microscopy and spectroscopy (STM-STS) data were acquired using a CreaTec low-temperature STM, utilizing an electrochemically etched tungsten tip. All the reported images and spectra have been collected either at 77 K or 4 K. For each STM image, the corresponding temperature is reported in the figure caption. Scanning Tunnelling Spectra were measured by first-derivative detection by means of an SRS lock-in amplifier: the bias voltage was modulated with a sinusoidal signal (amplitude 50 mV, frequency 2345 Hz, chosen to avoid any resonance with high harmonics of the power-line and other frequencies present in the system). The phase of the lock-in detection was determined by withdrawing the tip until no tunnel current was measured, so that only the capacitive component could flow, and selecting a value 90° smaller than the setting maximizing the capacitive signal. An integration time of 10 ms was chosen for the lock-in, and each spectrum was measured subsequently for an increasing and decreasing voltage ramp, lasting 30 s each, in order to detect any systematic effect. Depending on the noise level, 1-10 repeated measurements were averaged. Normalisation of the conductance units was carried out by comparison with the numerically calculated derivative of the current, which leads to a noisier, yet correctly normalized curve.

ARPES measurements, momentum maps and Photoemission Orbital Tomography (POT) of $WS_2$/Au(111) and 5A/$WS_2$/Au(111) systems were acquired with a KREIOS 150MM momentum microscope by SPECS GmbH. The instrument is coupled to two different light sources: a monochromatized helium discharge lamp (UVS300 and TMM150, from SPECS), and a high harmonic generation (HHG) chamber, generating photon energies of, respectively, 21.2 eV and 26.4 eV. The angle of incidence on the samples of both sources is 68° with respect to the surface normal. A detailed description of the HHG chamber can be found elsewhere[36]. Measurements were acquired both at 77 K and 6 K.

A 3D rendering of the full emission cone for the electronic band structure of $WS_2$/Au(111) is achieved using Blender equipped with the Microscopy Nodes[37] add-on.

**Computational.** For our theory approach, we simulated 5A on a monolayer of $WS_2$, ignoring the Au(111) substrate, since this would result in a prohibitively large unit cell due to the lattice mismatch between Au and $WS_2$. We use a supercell of [[5 −1][0 3]] (in terms of the $WS_2$ primitive cell with an in-plane lattice constant of 3.158 (Å) and place the molecular center above a sulfur atom. For the geometric structure optimizations within DFT, we use the QuantumEspresso code[38-40] with norm-conserving pseudopotentials from the PseudoDojo library[41,42] and the vdw-df-ob86 functional to account for exchange-correlation effects and van der Waals[43] interactions. The geometry relaxation was carried out until all forces were below 0.02 eV/Å, using a plane-wave cutoff of 80 Ry for the wave functions (320 Ry for the density) and a 2x4x1 shifted grid for the sampling of the Brillouin zone. To avoid spurious interaction between repeating slabs, we inserted a vacuum layer of 20 Å in conjunction with a truncation of the Coulomb interaction in z-direction[44]. Using the relaxed structure, we computed the charge density with the same settings as above, albeit with a 60 Ry cutoff for the plane waves, a k-mesh of 4x8x1, and the inclusion of spin-orbit interaction with a spinor representation of the wave functions. From the charge density, we compute wave functions and energies for the unoccupied states in a non-self-consistent run, using a $\overline{\Gamma}$-centered 6x12x1 k-mesh. Using the energies and wave functions from the DFT calculation, we computed the many-body corrections to the energies on the $G_0W_0$ level with the Yambo code[45,46]. For the screened interaction $W$, we used a sum over 2400 bands and a cutoff of 10 Ry, and 2400 bands for $G$ in conjunction with the Bruneval-Gonze method for accelerating convergence[47]. We truncate the Coulomb interaction in the direction perpendicular to the slab[48] and use a Brillouin zone (BZ) interpolation for 2D systems[49].

## RESULTS AND DISCUSSION

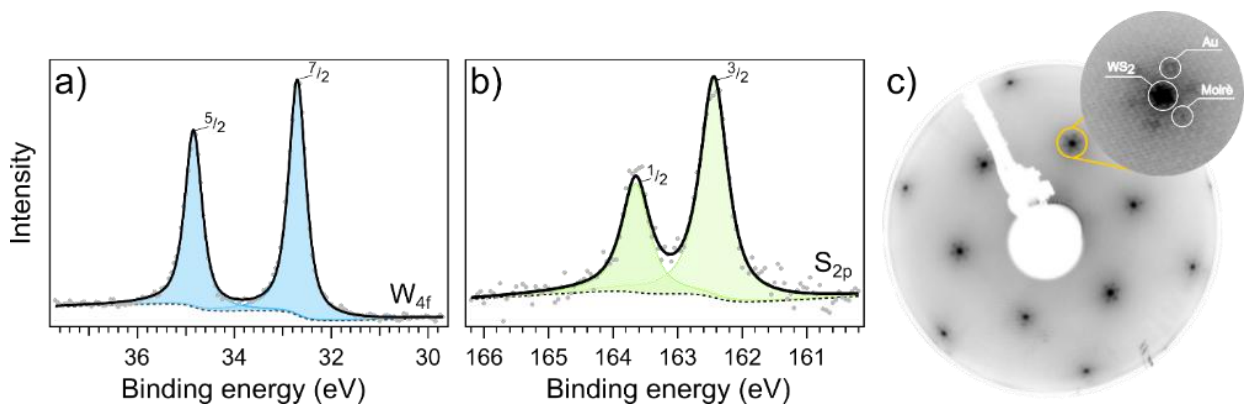

*Figure 1 – XPS spectra of monolayer $WS_2$ measured at room temperature. a) tungsten 4f orbitals and b) sulfur 2p orbital. c) LEED pattern of $WS_2$/Au(111) interface, showing the moiré pattern – inset - induced by slight mismatch in lattice constants for Au(111) and $WS_2$.*

First, we assess the chemical and crystalline quality of the $WS_2$ grown atop Au(111) from the DMDS precursor (more details can be found in the Methods section). The chemical composition of the as-deposited $WS_2$ monolayer has been checked by means of X-ray photoemission spectroscopy (XPS). Fig. 1 shows the acquired core level spectra for W $4f$ and S $2p$. The energy positions of $W4f_{7/2}$ and $S2p_{3/2}$ core levels (32.7 eV and 162.4 eV, respectively) are in excellent agreement with the reported values[50] for a monolayer of $WS_2$ grown on Au(111). Moreover, the lack of any detectable features at binding energies below 32.7 eV in Fig. 1a) proves the full sulphurization of tungsten atoms, excluding any significant sub-stoichiometric species or W clusters.

The surface quality and crystallinity have been investigated by means of low-energy electron diffraction (LEED) and STM. The LEED pattern of the as-grown $WS_2$ monolayer, shown in Fig. 1c), displays a clear moiré pattern, surrounding the main diffraction spot of $WS_2$. The inset in Fig. 1c) shows a close-up on the moiré pattern in which one of the outer points is superimposed to the gold LEED pattern[50]. This moiré pattern is typical of many TMDs grown on Au(111)[32,33,50] and it originates from the mismatch between the nominal lattice parameter[51] of Au(111), 2.89 Å, and the one of the TMD. Its presence also serves as an indicator of the high crystalline quality of the TMD and as a proof of epitaxial growth.

Figs 2a) and 2b) show, respectively, a large-scale and an atomically-resolved STM image of the $WS_2$ monolayer, providing an overview of its surface morphology. Two distinct features can be clearly identified in Fig. 2a): first, the moiré pattern also observed in the LEED measurements, and, second, a surface morphology characterized by randomly distributed depressions of varying sizes. These depressions, characterized by a height of a single Au(111) atomic step, are indicative of local structural inhomogeneities or possible growth-related defects.

From an STM line-scan across a step edge (Fig. 2c), we determine a step height of 0.2 nm, which is consistent with Au(111) atomic steps and thereby confirms the continuity of the $WS_2$ film across the gold surface, see also Fig. S1. The depression pattern which characterizes the entire surface has been already observed on gold surfaces exposed to high doses of sulfur[52]: due to its high reactivity, sulfur corrodes the gold surface, yielding one-atom deep depressions (see line profile in fig. 2c), where the depth of the depression is seen to match the height of a substrate step edge). Nevertheless, we observe a continuous moiré pattern across boundaries of those depressions, meaning that WS₂ forms a seamless film covering/carpeting the depressions as well as step edges, as also seen in Fig. S1.

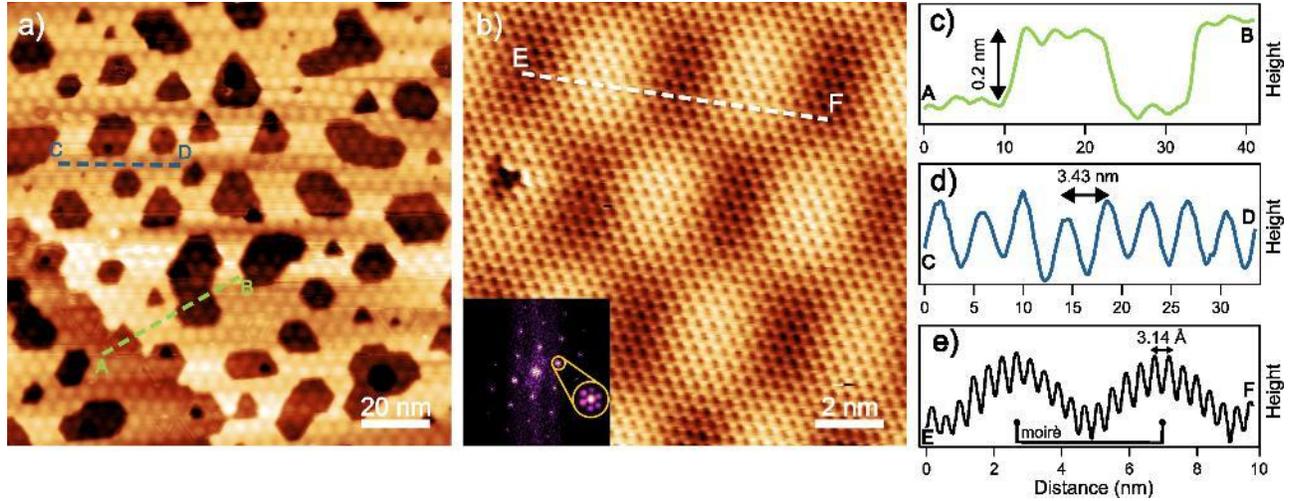

*Figure 2 – a) Large scale STM image of WS₂/Au(111) interface; $V = 0.084V, I = 20\,pA, (130 \times 130)\,nm^2, T = 77K$; b) Atomic resolution of WS₂ monolayer; $V = -1.4\,V, I = 20\,pA, (13 \times 13\,)nm^2, T = 4K$; inset shows the Fast Fourier Trasform of image b) where moirè satellite spots are observed, as in the LEED pattern; c) Height profile across Au(111) terrace step and hole,, consistent with the height of single layer of Au(111); d) line profile showing the moirè periodicity; e) line profile showing the lattice constant of WS₂ superimposed to the moirè periodicity induce by the underneath Au(111).*

The measured periodicity associated with the moiré pattern is 3.43 nm, Fig. 2d), while in Fig. 2e) we report the line-scan from the atomically-resolved image of the WS₂ monolayer, Fig. 2b), where it is possible to observe the moiré pattern superimposed on the atomic unit cell of WS₂, measured as 0.314 nm and consistent with the lattice constant reported in literature[53]. These parameters are also consistent with the moiré empirical formula $d_{moirè} = \frac{a_{Au}a_{WS_2}}{|a_{Au} - a_{WS_2}|} = 3.48\,nm$.

The measured values for the WS₂/Au(111) lattice constant and moiré periodicity of, respectively, 0.316 nm and 3.2 nm, reported in a previous work[53] are in fair agreement with our values of 0.314 nm and 3.43 nm. Note that the slight differences also result in different commensurate moiré supercells. Our results suggest $(11 \times 11)$ for WS₂ and $(12 \times 12)$ for Au(111), while the previously reported moiré supercell for WS₂ and Au(111) were $(10 \times 10)$ and $(11 \times 11)$, respectively.

We now discuss the electronic properties of the WS₂/Au(111) interface, investigated by means of both ARPES and scanning tunneling spectroscopy (STS). In Fig. 3a), we show the electronic band dispersion for the WS₂ monolayer along the high symmetry directions indicated in Fig. 3b). Clear spin-split bands at the high symmetry point $\bar{K}$, characteristic of both W and Mo based TMDs, confirm the high quality of the grown WS₂ monolayer, with a measured band splitting of 417 meV, consistent with previously reported values[54,55]. Since the growth of WS₂ on metals does not follow a self-terminating protocol, the second layer is expected to nucleate[56] already from coverages of 0.7 ML.

As a direct-to-indirect bandgap transition is observed on bilayer formation,[1,10,11] assessing the number of layers is crucial for understanding the band alignment at the interface. Our ARPES data taken with a photon energy of 21.2 eV show that the valence band maximum (VBM) at the $\overline{K}$ point is higher (direct band gap) than at $\overline{\Gamma}$ (see Fig. 3a), suggesting that we are in the monolayer regime[56]. To exclude the presence of other electronic bands that could be suppressed by final states effects, we also acquired ARPES spectra at higher photon energy of 40.2 eV (as reported in Fig. S2). These data, shown in Fig. S2d), show the presence of an additional band at $\overline{\Gamma}$, albeit extremely faint, which could be related to the presence of the second layer. Despite this evidence, no STM images with a clear presence of the second layer have been measured. Moreover, from the same dataset, we observed extra bands at the $\overline{M}$ point, which are consistent with the formation of projected band gap edges induced by the interaction with the underlying Au(111) substrates[57].

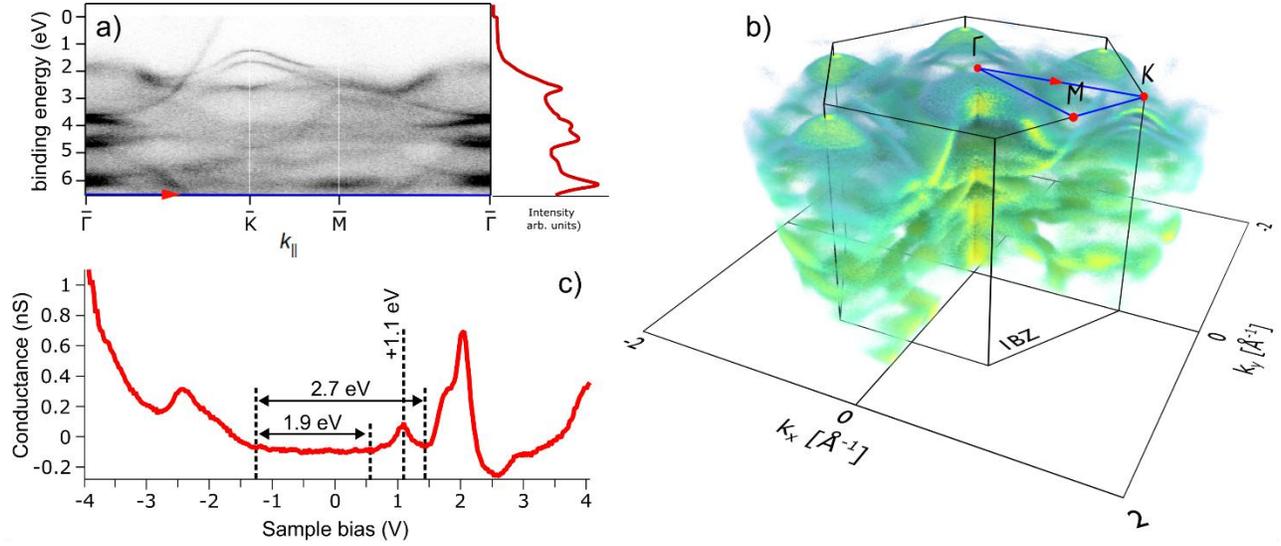

*Figure 3 – a) ARPES spectra of monolayer WS₂/Au(111) along high symmetry directions in the Brillouine zone (shown in b). The characterstic spin split bands of WS₂ are clearly visible at the **K** point, while the inset on the right shows the angle integrated spectrum of the WS₂/Au(111) interface; $\hbar\omega = 21.2\ eV, T = 6K$; b) 3D rendiring of the full emission cone/bands of WS₂/Au(111), $\hbar\omega = 21.2\ eV, T = 6K$; c) dI/dV spectrum acquired on monolayer WS₂/Au(111), showing a 2eV gap and interface states located at +1eV, set point $\Delta V = 4\ V, I = 30\ pA, T = 77K$;*

Figure 3c) shows the dI/dV spectrum acquired on pristine WS₂/Au(111) interface. We first discuss the peak located at +1.1 V, i.e. in the unoccupied states. This interface state has already been reported for MoS₂ and WS₂ grown on Au(111)[58,59], and it has been attributed to a hybrid state arising from the interaction between the TMD and Au(111). However, further investigation of the nature of this state is beyond the scope of the present study. We also point out that the presence of a negative conductance at $\sim +2.5\ V$ is an artifact introduced by the low-noise transimpedance amplifier (TIA) of the STM caused by a steep and fast variation of the signal in that point.

The evaluation of the transport gap size is complicated by both the presence of the interface state and the fact that the tunneling process is strongly suppressed for electrons with momentum[58] $k \neq 0$, i.e. away from $\overline{\Gamma}$. Regarding the former, if we would evaluate the band gap from the valence band maximum at -1.3 V to the onset of the interface state at +0.6 V, this would lead to a bandgap of around 1.9 eV, which is somewhat smaller than the reported range of 2.2 – 2.7 eV for freestanding WS₂ monolayer[60,59,61]. Instead, if the gap is evaluated using the feature at higher positive bias voltage, this would lead to a band gap of around 2.7 eV. The second issue, namely the k-dependence of the tunneling probability, is also particularly relevant, as the direct band gap of the monolayer $WS_2$ lies at the $\overline{K}$ point. This introduces an uncertainty in the assignment of the valence band maximum and conduction band minimum (CBM) via STS, which will also become important in the second part of the paper, where the energy level alignment of pentacene/$WS_2$ is discussed.

The band gap derived from STS is in good agreement with our $G_0W_0$ calculations for a free standing $WS_2$ layer where we obtain a transport gap of 2.5 eV. To summarize, we consider the interface state to lie within the $WS_2$ transport gap, and, taking into account the k-dependence of the tunneling signal and the error it introduces, we set the experimental upper limit of the transport gap to 2.7 eV.

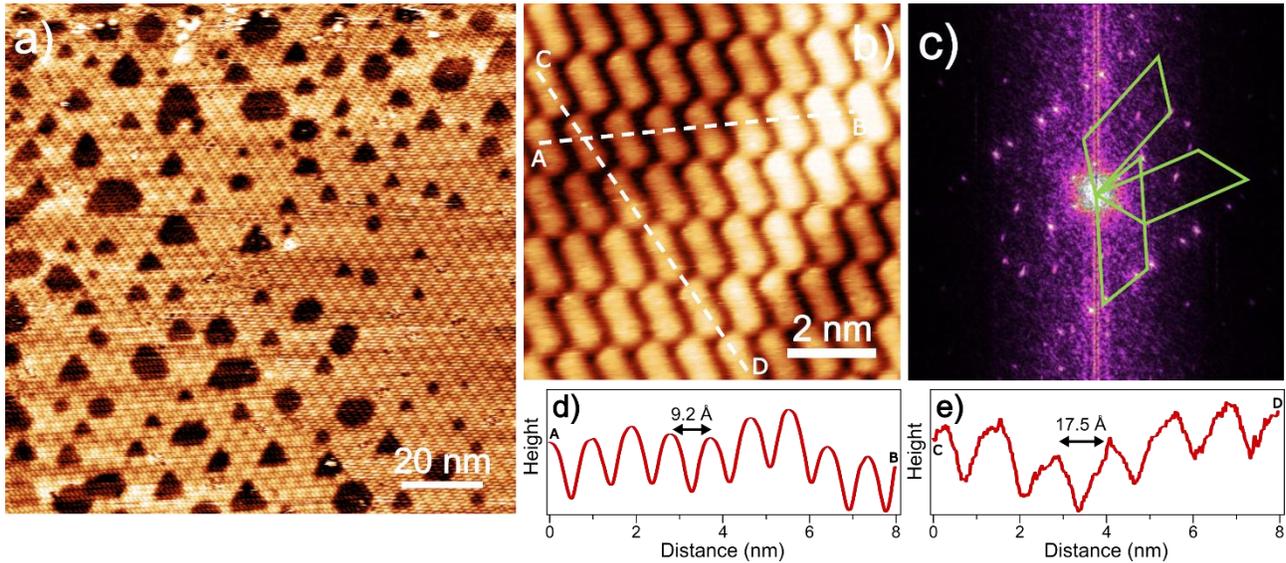

*Figure 4 - a) Large scale STM image of pentacene self assembly on $WS_2$/Au(111) interface, $V = -1.4\ V, I = 20\ pA, (130 \times 130)\ nm^2, T = 77K;$ b) Close up of pentacene molecular layer atop $WS_2$ monolayer, $V = -1.4\ V, I = 20\ pA, (8.5 \times 8.5)\ nm^2, T = 77K;$ c) Fast Fourier Transform of image a) where the three different rotational domains of 5A on the hexagonal symmetry of $WS_2$ are highlighted in green; d) measured periodicity of pentacene self-assembled monolayer short axis; e) measured periodicity of pentacene self-assembled monolayer long axis.*

We next address how this system is modified upon molecular adsorption of pentacene, focussing first on the geometrical properties of the adsorbate layer. The deposition of a single layer of pentacene molecules on pristine $WS_2$ results in a self-assembled structure characterized by long-range ordering, as confirmed by the large-scale STM image shown in Fig. 4a). Note that, at the chosen bias, the $WS_2$/Au(111) moiré pattern is still visible through the molecular layer. The self-assembled molecular structure is identified by high resolution STM images presented in Fig. 4b) and by Fourier analysis of the large scale STM image in Fig. 4c). Both reveal that pentacene forms an oblique unit cell with, respectively, 0.92 nm, Fig. 4d), and 1.75 nm, Fig. 4e), lattice constants, leading to a commensurate $p(3, \sqrt{21})$ superstructure with respect to $WS_2$. Due to the hexagonal symmetry of $WS_2$, 3 rotational domains are observed, as elucidated in Fig. 4c). It is worth noting that the continuity of the grown $WS_2$ monolayer above the one-atom deep Au depressions is also reflected by the molecular super-structure, which does not show evidence of accumulation points in presence of the depression edges, as could be expected in the proximity of atomic steps.

Having established the structural properties of the 5A/WS$_2$ hybrid interface, we next focus on its electronic properties, which we first investigate by means of STS. The corresponding spectra are shown in Fig. 5a. The dI/dV curve acquired on the 5A/WS$_2$ interface is superimposed on the one acquired on pristine WS$_2$/Au(111) (in red).

The STS of the 5A/WS$_2$ interface has two prominent peaks located at -1.3 eV and +1.8 eV. We were able to achieve orbital resolution at the corresponding bias voltages, Fig. 5b and 5c, which can be unambiguously associated with the HOMO and LUMO, respectively [62], revealing a HOMO-LUMO gap of 3.1 eV. The magnitude of the pentacene gap also acts as an indication of its interaction with the underlying WS$_2$ substrates: for gas-phase pentacene (non-interacting) the reported [62] gap is 5.2 eV, while it shrinks to 2.2 eV for 5A on Au(111)[63]. In between this range, other reported 5A gaps on non-metallic surfaces are 3.4 eV for 5A/h-BN/Rh(111)[64] and 4.1 for 5A/NaCl/Cu(111)[65].

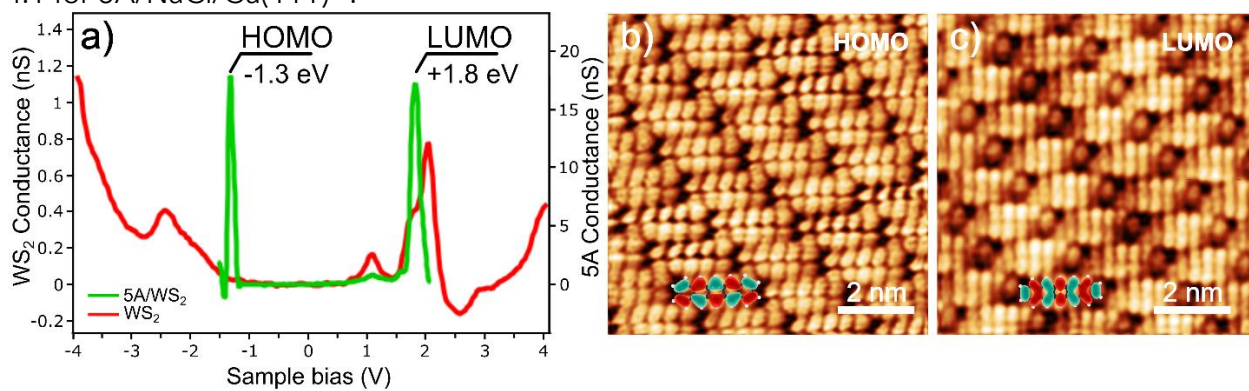

*Figure 5 – a) dI/dV spectrum of pentacene monolayer atop WS$_2$/Au(111), green line, superimposed to the dI/dV spectrum of pristine WS$_2$/Au(111), red line, set point; b) STM image of 5A-HOMO molecular orbital, $V = -1.3\,V$, $I = 20\,pA$, $(7 \times 7)\,nm^2$, $T = 77\,K$; c) STM image of 5A-LUMO molecular orbital, $V = +1.8\,V$, $I = 20\,pA$, $(7 \times 7)\,nm^2$, $T = 77\,K$,*

Concerning the electronic level alignment, we observe that the HOMO lies slightly above the valence band maximum, i.e. within the electronic gap of WS$_2$, while the LUMO is located within the WS$_2$ conduction band. The reliability of our STS measurement is further confirmed by the presence of the hybrid states at +1 eV also in the spectra from the 5A/WS$_2$/Au(111) interface, ensuring that no charging effects played a role in artificially shifting the HOMO-LUMO peaks positions.

Despite the small energy difference between the VBM and HOMO, and the uncertainty in determining the exact VBM position, as discussed above, the energy level alignment suggests a type-II band alignment.

In order to corroborate these experimental findings, we perform electronic structure calculations with the experimentally obtained geometry as a starting point, albeit without the underlying Au(111) substrate. We use van der Waals-corrected density functional theory (DFT) to relax the geometry and obtain the structure shown in Figure 6 a). Since the correct energy level alignment in such hybrid systems is far from trivial and DFT is known to typically underestimate the gap, we employ many-body perturbation theory at the $G_0W_0$ level on top of the DFT results. While DFT with a semi-local exchange-correlation functional would incorrectly predict a type-I level alignment, our $G_0W_0$ results present a considerable improvement over DFT. Direct comparison between the obtained STS spectra (see Fig. 5a) and the projected density of states from $G_0W_0$ (Fig. 6b), shows a fair agreement, both for the molecular and the $WS_2$ features.

On a quantitative level, the theoretical results deviate slightly from the experiment. For instance, we observe that the calculated HOMO-LUMO gap is somewhat smaller than in experiment (2.8 eV vs. 3.1 eV), which is a known effect of the starting point for $G_0W_0$ calculations and could be mitigated by self-consistent GW. For a system of this size, the latter is computationally very demanding, and beyond the scope of this work.

Given the challenges in both the computational and experimental approaches presented so far, a complementary technique is expedient in order to gain additional insight into the energy alignment, especially around the VBM. In this regard, photoemission orbital tomography (POT)[66] enables the unambiguous identification of molecular orbitals of organic adsorbates and their energy positions, as well as the valence band signatures of $WS_2$.

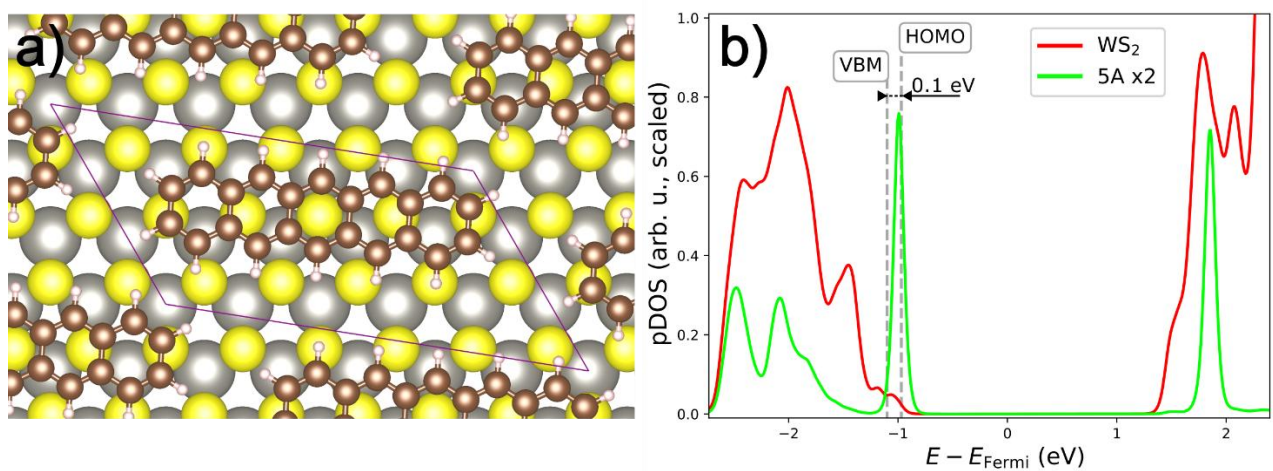

*Figure 5. - Structure used in the DFT/$G_0W_0$ calculations. W atoms are depicted in silver, S in yellow, C in brown and H in white. The unit cell is indicated in purple. (b) Density of states from $G_0W_0$ projected on $WS_2$ (green) and on 5A (red, scaled by a factor of 2). Dashed lines marked with HOMO and VBM show the energy positions for the momentum maps depicted in Fig. 7 d) and e), respectively.*

Within the POT approach, we measure the momentum distribution of the photoemitted electrons from a molecular film at a given kinetic energy, and we compare it with a cut of the 3D Fourier transform of a given molecular orbital. In order to identify the molecular resonances of interest at first, we compare the momentum-integrated photoemission spectra for the $WS_2$/Au(111) interface before and after the deposition of pentacene. The corresponding spectra are reported in Figure 7a). To better elucidate the changes induced by the molecular adlayer, the difference between the two spectra is also shown. We determined the energy position of the molecular features by fitting the molecular peaks with a Gaussian function.

Upon molecular adsorption, two new peaks located at energies of -0.8 eV and -1.1 eV emerge. While the former (FWHM: 0.20 eV) is associated with 5A in direct contact with Au(111), (see also Fig. S4), and is present because of the incomplete coverage (0.7 ML) of the Au(111) surface with $WS_2$, the peak at -1.1 eV (FWHM: 0.17 eV) arises from 5A on $WS_2$. The energy shift between the two pentacene species is due to the interaction with the different surface underneath. In order to assign the peak originating from 5A on $WS_2$ to a specific 5A orbital, within the POT approach, we simulate photoemission momentum maps using a plane-wave final state[67] and the DFT wave functions in conjunction with $G_0W_0$ energy levels. This allows us to compare the experimental momentum map taken at the 5A HOMO energy position, Fig. 7 b), with the simulated map, Fig. d). We would like to highlight that, since POT is a space-average technique, we need to take into account the different rotational domains of the substrate in order to obtain a meaningful comparison between theory and experiment. Therefore, we apply the corresponding symmetry operations to our simulated momentum maps. The overall procedure is explained in detail elsewhere[68]. Our comparison between the simulated and measured momentum maps, Fig. 7 b) and d), respectively, confirms that the peak at -1.1 eV is indeed the Pentacene HOMO. The round feature at $\overline{\Gamma}$ originates from the $WS_2$ second layer and becomes visible here because of the higher photon energy employed (namely 26.4 eV). It is worth noticing that even if the absolute values measured with different techniques are slightly different due to the different physical processes involved in each measurement, the relative positions between 5A HOMO and $WS_2$ VBM are always consistent, i.e. the HOMO is always located at higher energy than $WS_2$ VBM, thus proving the type-II energy level alignment.

We would also like to point out that the apparent energy degeneracy of the HOMO feature and the $WS_2$ VBM in the calculations, as reported in the projected DOS presented in Fig. 6 b), is somehow relaxed when simulating momentum maps. This is because the spectral weight of the sixfold features of the $WS_2$ VBM is so low that the feature at this energy is dominated by emission from the HOMO level.

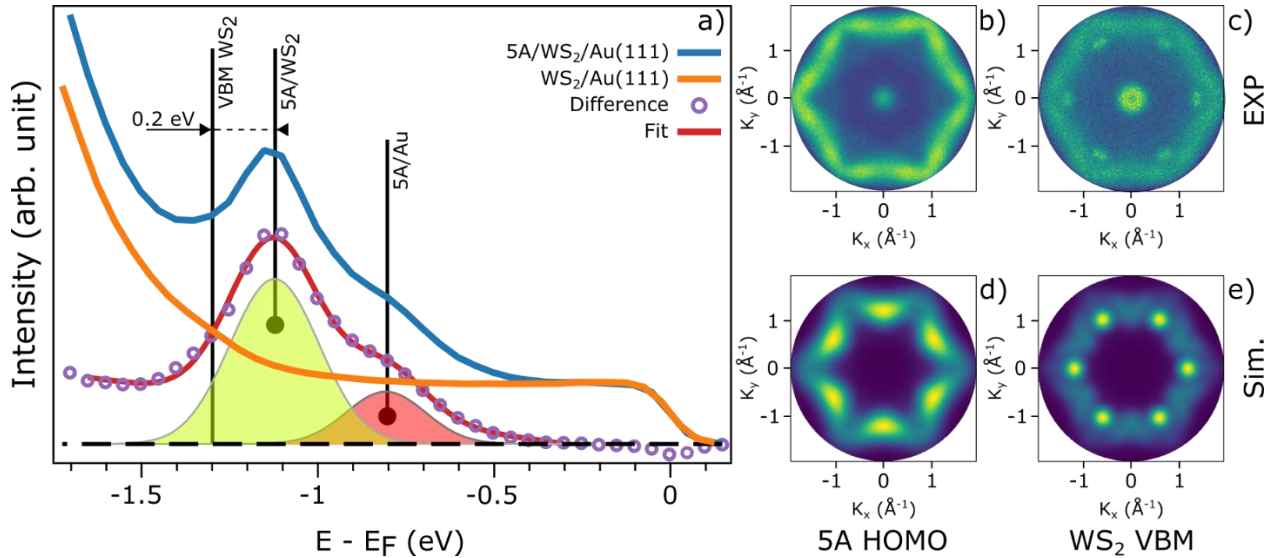

*Figure 7 – a) Comparison of UPS spectra: pristine $WS_2$/Au(111), orange line, and pentacene monolayer, blue line. To enhance pentacene features, the difference between the UPS spectra, before and after 5A deposition, is plotted, red line; b)-c) Symmetrized experimental pentacene POT averaged in the energy range associated only to 5A/$WS_2$ interface and VBM of pristine $WS_2$/Au(111); d)-e) DFT calculated 5A/$WS_2$/Au(111) POT and $WS_2$/Au(111) VBM. All spectrum and k-maps are acquired at $\hbar\omega = 26.4\ eV, T = 77K$.*

We now compare an experimental momentum map measured at an energy of -1.3 eV, 200 meV below the HOMO feature (Fig. 7c), with the simulated map taken 100 meV below the HOMO feature in the calculated DOS (Fig. 7e). These energies correspond to the onset of the VBM. Also in this case, we find good agreement between theory and experiment: the six-fold features, which are characteristic of the $WS_2$ valence band maximum can be easily recognized in both the experimental and the simulated momentum maps. Another striking similarity between the two datasets is the presence of a faint background feature associated with the momentum-resolved HOMO signal. For the experimental data, this can be attributed to the large FWHM of the 5A HOMO emission, as shown in Fig. 7a). Here, the left tail of the 5A HOMO peak is slightly superimposed to the $WS_2$ VBM, located ~0.2 eV below the pentacene feature, but we emphasize that the overall spectral weight is quite low. For our simulations, a similar argument holds.

Our overall conclusion is that the interface exhibits a type-II energy level alignment, supported by STS data and further corroborated by POT and theoretical calculations.

## CONCLUSION

In summary, we have grown a novel hybrid organic-TMD interface, pentacene on $WS_2$, and employed a complementary spectro-microscopy approach to investigate its structural and electronic properties. By means of molecular beam epitaxy growth, we were able to obtain a high quality, extended and atomically flat $WS_2$ film, using a novel precursor, namely DMDS, as a sulfur source.

The direct growth of TMD on metallic single crystal substrates enables single crystalline domains, but yields strong interfacial coupling, as observed by the presence of an interface hybrid state in STS spectra. Subsequent deposition of a single layer of pentacene results in an ordered commensurate superstructure on top of $WS_2$, which is characterized by three distinct rotational domains and an oblique unit cell. STS spectra, supported by $G_0W_0$ calculations, locate the pentacene LUMO within the conduction band of the $WS_2$, while being undecisive in determining the precise relative energy position of the HOMO with respect to the valence band maximum. This ambiguity has been resolved by our POT data which revealed that the HOMO lies above the $WS_2$ VBM, indicating the energy level alignment to be of a type-II.

This system represents therefore an ideal starting point to achieve a separation of electron and holes in different layers. More in detail, we suggest that the possibility of intercalating a decoupling layer at the $WS_2$/Au(111) interface might be an effective method to prevent the quenching of excitonic resonances. In fact, this particular energy level alignment, together with an extended and well-ordered interface, may favor the establishment of interlayer excitons between $WS_2$ and pentacene, making this interface a prototypical system for studying non-equilibrium phenomena with time-resolved based techniques.

**Acknowledgements:**

G.Z., M. Capra and S.T. acknowledge the support from the Deutsche Forschungsgemeinschaft (DFG, German Research Foundation), Project ID 524569125. IDB acknowledges support from the Ramon y Cajal program, grant no. YC2022-035562-I, and financial support to the IFIMAC foundation from the Spanish Ministry of Science and Innovation through the María de Maeztu" Programme for Units of Excellence in R&D (Grant CEX2023-001316-M).

C.S.K and P.P acknowledge support from the European Research council (ERC) Synergy Grant, Project ID 101071259 and the EuroHPC JU for awarding the project ID EHPC-EXT-2024E02-054 access to Leonardo at CINECA, Italy.